\documentclass[sigconf,nonacm]{acmart}

\usepackage{booktabs}
\usepackage{multirow}
\usepackage{graphicx}
\usepackage{xcolor}
\usepackage{balance}
\usepackage{float}

\graphicspath{{figures/}}

\title{Less is More: Benchmarking LLM Based Recommendation Agents}

\author{Kargi Chauhan}
\authornotemark[1]
\affiliation{%
  \institution{University of California, Santa Cruz}
  \country{USA}
}
\email{kchauha3@ucsc.edu}

\author{Mahalakshmi Venkateswarlu}
\authornote{Both authors contributed equally to this research.}
\affiliation{%
  \institution{Georgia Institute of Technology}
  \country{USA}
}
\email{mvenkate3@gatech.edu}

\begin{abstract}
Large Language Models (LLMs) are increasingly deployed for personalized product recommendations, with practitioners commonly assuming that longer user purchase histories lead to better predictions. We challenge this assumption through a systematic benchmark of four state of the art LLMs GPT-4o-mini, DeepSeek-V3, Qwen2.5-72B, and Gemini 2.5 Flash across context lengths ranging from 5 to 50 items using the REGEN dataset.

Surprisingly, our experiments with 50 users in a within subject design reveal no significant quality improvement with increased context length. Quality scores remain flat across all conditions (0.17--0.23). Our findings have significant practical implications: practitioners can reduce inference costs by approximately 88\% by using context (5--10 items) instead of longer histories (50 items), without sacrificing recommendation quality. We also analyze latency patterns across providers and find model specific behaviors that inform deployment decisions. This work challenges the existing ``more context is better'' paradigm and provides actionable guidelines for cost effective LLM based recommendation systems.
\end{abstract}


\begin{document}

\maketitle


\section{Introduction}

Large Language Models (LLMs) have emerged as powerful tools for personalized recommendations, capable of understanding user preferences from purchase histories and generating contextually relevant product suggestions~\cite{fan2023recommender, wu2024survey}. Unlike traditional collaborative filtering approaches that rely on learned embeddings, LLM-based recommenders can directly process rich textual descriptions of user history, item metadata, and contextual information. As recommendation systems evolve into agentic ecosystems capable of planning, reasoning, and multi-turn interaction, the ability to efficiently manage context becomes foundational for scalable and responsive personalization~\cite{wang2023recmind, zhang2024agent4rec}.

A common assumption among practitioners is that providing more user history longer context leads to better recommendation quality. This belief drives the use of extended context windows, often including dozens or even hundreds of past interactions. However, this assumption remains largely untested in recommendation scenarios. While LLMs have demonstrated impressive capabilities in utilizing long contexts for tasks like document summarization and question answering, their ability to leverage extended user histories for product recommendations is unclear~\cite{liu2024lost}. Moreover, longer contexts come with significant costs: increased API expenses, higher latency, and greater computational requirements.

This question has critical implications for deploying recommendation agents. More context may help the LLM better understand user preferences, but excessive history could introduce noise or cause the model to lose focus on recent, relevant interactions~\cite{quadrana2018sequence}. Longer prompts require more computation, directly impacting response times in real time recommendation scenarios. API based LLM services typically charge per token, making context length a direct cost driver for large scale deployments. Additionally, LLM context windows are a finite resource; identifying the minimal sufficient history enables autonomous agents to preserve capacity for multi turn reasoning and long horizon planning~\cite{wang2023recmind}.

In this paper, we systematically investigate the relationship between context length and recommendation quality across four state of the art LLMs. Our research question is straightforward yet critical:
\begin{quote}
\textit{Does providing more user purchase history improve LLM based product recommendation quality?}
\end{quote}

To answer this question, we conduct controlled experiments using the REGEN dataset~\cite{su2025regen}, testing context lengths of 5, 10, 15, 25, and 50 items across GPT-4o-mini, DeepSeek-V3, Qwen2.5-72B, and Gemini 2.5 Flash. We employ a within subject design where the same 50 users are evaluated at all context lengths, ensuring fair comparisons. Our key contributions and findings are:
\begin{enumerate}
    \item \textbf{Systematic Benchmark}: We design and execute experiments measuring recommendation quality and latency across multiple context lengths using four production LLMs from different providers.
    \item \textbf{Flat Quality Curves}: All four models show no significant quality improvement as context length increases from 5 to 50 items. Quality scores remain in the 0.17--0.23 range with overlapping confidence intervals.
    \item \textbf{Universal Pattern}: This finding holds across models from four different providers (OpenAI, Google, Alibaba, DeepSeek), suggesting a fundamental limitation rather than model-specific behavior.
    \item \textbf{Cost Implications}: Token usage increases approximately 8$\times$ when moving from 5 to 50 items, while quality remains unchanged. Practitioners can save up to 88\% on inference costs by using minimal context.
    \item \textbf{Latency Analysis}: We identify model specific latency patterns that inform deployment decisions for real time recommendation scenarios.
\end{enumerate}

These findings challenge the ``more context is better'' paradigm and provide evidence based guidelines for deploying cost effective LLM based recommendation systems. As LLM based agents become central to recommendation ecosystems, understanding their computational efficiency is critical for scalable deployment. Our results show that LLM-based recommendation systems work just as well with 5--10 items of user history as with 50 items, enabling significant cost savings and faster response times for large scale deployments.


\section{Related Work}

\subsection{LLMs for Recommendations}

The integration of LLMs into recommendation systems has gained significant attention in recent years. Fan et al.~\cite{fan2023recommender} provide a comprehensive survey of LLM based recommender systems, highlighting their potential for understanding user intent and generating explanations. P5~\cite{geng2022p5} unified multiple recommendation tasks through a text to text framework, demonstrating that a single language model can handle rating prediction, sequential recommendation, and explanation generation. TALLRec~\cite{bao2023tallrec} showed that LLMs can be effectively tuned for recommendation with limited training data, achieving competitive performance against specialized models. Hou et al.~\cite{hou2024large} demonstrate that LLMs can achieve competitive performance on sequential recommendation tasks when properly prompted.

More recently, agent based approaches have emerged.     RecMind~\cite{wang2023recmind} frames recommendation as a planning problem where an LLM agent reasons over user preferences. Agent4Rec~\cite{zhang2024agent4rec} simulates user recommender interactions to evaluate recommendation strategies. Recent work has also explored conversational recommendation systems where LLMs engage in multi-turn dialogues to elicit user preferences~\cite{friedman2023leveraging}. InstructRec~\cite{zhang2023instructrec} demonstrated that instruction tuned LLMs can follow natural language preference descriptions to generate recommendations. Additionally, Chat-Rec~\cite{gao2023chatrec} proposed augmenting LLMs with traditional recommender systems to combine the strengths of both paradigms. However, most existing work focuses on demonstrating LLM capabilities rather than systematically studying the factors that influence recommendation quality. Our work fills this gap by specifically investigating the role of context length.

\subsection{Why Doesn't More Context Help?}

Our findings challenge the intuitive assumption that more user 
history should lead to better recommendations. Several factors 
may explain this pattern.

First, the \textit{Lost in the Middle} phenomenon identified by 
Liu et al.~\cite{liu2024lost} suggests that LLMs struggle to 
effectively utilize information positioned in the middle of long 
contexts. When context grows from 5 to 50 items, most information 
falls into this underutilized region.

Second, \textit{recency bias} may render older purchases less 
predictive of future behavior. The most recent 5 items likely 
capture a user's current interests as effectively as 50 items, 
consistent with patterns observed in sequential recommendation 
research~\cite{quadrana2018sequence}.

Third, \textit{signal saturation} implies diminishing returns to 
additional context. The first few items may sufficiently establish 
user preferences, with additional items contributing noise rather 
than signal.

Finally, \textit{task difficulty} imposes a ceiling on performance. 
Predicting the exact next product purchase is inherently challenging; 
even perfect context utilization may not yield substantial quality 
improvements.

\subsection{Sequential Recommendation}

Traditional sequential recommendation methods like SASRec~\cite{kang2018sasrec} and BERT4Rec~\cite{sun2019bert4rec} use self-attention to model user behavior sequences. These models typically process fixed length histories (e.g., 50--200 items) and have been extensively benchmarked on optimal sequence lengths. Our work extends this analysis to LLM based recommenders, which operate under different computational constraints.

\subsection{Efficient Recommendation}

Efficiency in recommendation systems has traditionally focused on model compression, approximate nearest neighbor search, and caching strategies~\cite{zhang2023efficiency}. For LLM-based systems, new optimization opportunities arise. LLMLingua~\cite{jiang2023llmlingua} proposes prompt compression techniques that reduce token count while preserving semantic content. Selective Context~\cite{li2023compressing} uses self-information to identify and remove less informative tokens. Systematic benchmarking has been valuable in understanding LLM capabilities and limitations, as demonstrated by studies like HELM~\cite{liang2022holistic}. Our work contributes a focused benchmark specifically examining context length effects in recommendation scenarios.


\section{Methodology}

\subsection{Problem Formulation}

Given a user $u$ with interaction history $H_u = \{i_1, i_2, \ldots, i_n\}$ ordered chronologically, and a target item $i_{\text{target}}$ that the user will interact with next, we study how the length $k$ of the history subset $H^k_u = \{i_{n-k+1}, \ldots, i_n\}$ affects: (1) Quality: The accuracy of predicting $i_{\text{target}}$; (2) Latency: The time to generate a recommendation; and (3) Token Usage: The computational cost (proxy for API cost). We evaluate these metrics across four different LLM providers to understand model specific behaviors.

\subsection{Dataset}

We use the REGEN (Reviews Enhanced with GEnerative Narratives) dataset~\cite{su2025regen}, which extends Amazon Product Reviews with rich annotations. The dataset includes, Purchase History which is sequential user item interactions with item metadata (title, category, price) and review information (rating, text), Ground Truth is the actual next purchase with explanatory narratives describing why the user made that choice, and User Profiles are summarized user preferences derived from their interaction patterns.

We focus on the Office Products category, which provides users with substantial purchase histories. From this dataset, we select users with at least 51 items in their purchase history (50 items for context + 1 item as ground truth for evaluation). Table~\ref{tab:dataset} summarizes the dataset statistics.

\begin{table}[t]
\centering
\caption{Dataset Statistics for Office Products Domain}
\label{tab:dataset}
\smallskip
\begin{tabular}{@{}lc@{}}
\toprule
\textbf{Statistic} & \textbf{Value} \\
\midrule
Total Users & 89,489 \\
Avg. History Length & 7.9 \\
Min History Length & 5 \\
Max History Length & 303 \\
Users with 51+ items & 189 \\
Users with 25+ items & 1,468 \\
\bottomrule
\end{tabular}
\end{table}

\subsection{Models}

We evaluate four state of the art LLMs representing different providers and architectures, GPT-4o-mini (OpenAI) a cost-efficient model from the GPT-4 family, accessed via the OpenAI API. DeepSeek-V3 (DeepSeek) an open-weight model with strong performance, accessed via the Together AI API. Qwen2.5-72B (Alibaba) a large scale multilingual model, accessed via the Together AI API and Gemini 2.5 Flash (Google) a fast inference model optimized for efficiency, accessed via the Google AI API.

\subsection{Experimental Design}

We employ a within subject design where the same 50 users are tested at all context lengths. This design controls for user-level variability and enables fair comparisons across conditions. Users are randomly sampled with a fixed seed (42) for reproducibility.

We test five context lengths: 5, 10, 15, 25, and 50 items. For each user, we use the most recent $k$ items as context and predict the next purchase, always taking the most recent interactions to preserve recency information.

Each model receives a standardized prompt containing the user's purchase history (title, category, rating) and is asked to predict the next product purchase with reasoning:
\begin{verbatim}
Based on this user's purchase history,
predict what product they will buy next.

USER PURCHASE HISTORY (most recent last):
1. [Item Title]
   Category: [Category]
   Rating: [X]/5
...
N. [Item Title]
   Category: [Category]
   Rating: [X]/5

What product will this user purchase next?
\end{verbatim}

\subsection{Evaluation Metrics}

We use a composite quality score combining keyword overlap and category matching:
\begin{equation}
\text{Quality} = 0.7 \times \text{KeywordScore} + 0.3 \times \text{CategoryMatch}
\end{equation}

The 0.7/0.3 weighting prioritizes semantic relevance (keyword overlap) 
over categorical accuracy, reflecting the intuition that predicting 
the right type of product within a category is more valuable than 
merely matching the broad category. This weighting aligns with 
evaluation practices in prior zero shot recommendation work~\cite{hou2024large}, 
where lexical similarity metrics are given higher weight than 
coarse grained category matches.
where KeywordScore is the Jaccard like overlap between predicted and ground truth product keywords (excluding stopwords), and CategoryMatch is a binary indicator (1.0 if category appears in prediction, 0.0 otherwise). This metric captures both semantic relevance (keyword overlap) and categorical accuracy (category match), similar to evaluation methods used in prior work~\cite{hou2024large}. While our composite Quality Score provides a reliable semantic and categorical proxy for zero shot performance, we acknowledge that complex agentic interactions may require metrics that capture long horizon impact. Future extensions will incorporate sophisticated ranking metrics such as Mean Reciprocal Rank (MRR) and Normalized Discounted Cumulative Gain (NDCG) to better evaluate performance across diverse item spaces.

Latency is Wall clock time from API request initiation to response receipt, measured in seconds.

Token Count is the number of input tokens as reported by each API, serving as a proxy for computational cost.


\section{Results}

\subsection{Main Finding: Flat Quality Curves}

\begin{figure}[!t]
    \centering
    \includegraphics[width=\columnwidth]{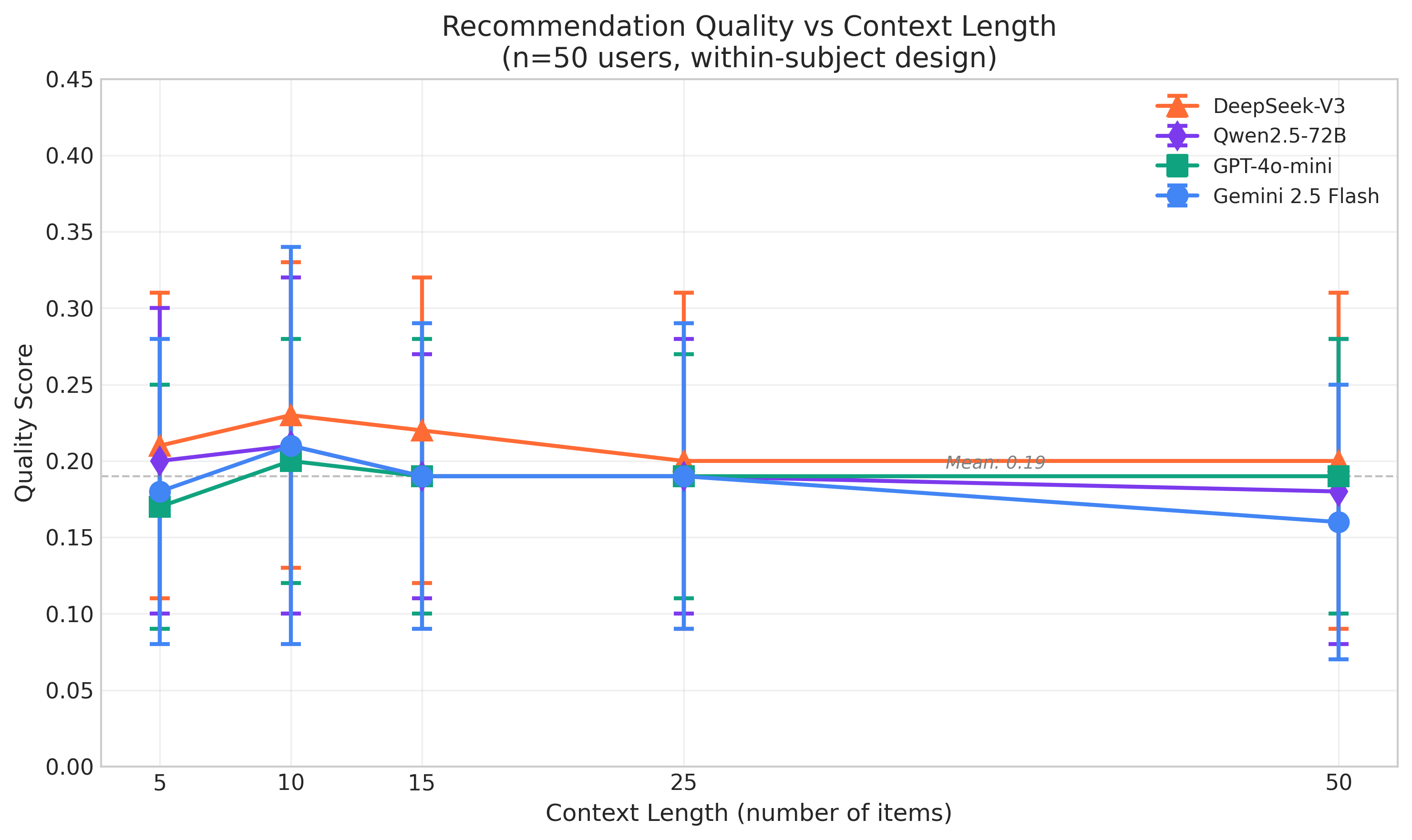}
    \caption{Recommendation quality vs. context length across four LLMs (n=50 users). All models show flat quality curves with overlapping confidence intervals, indicating no significant improvement with longer context.}
    \label{fig:quality}
\end{figure}

Table~\ref{tab:main_results} presents our complete results across all models and context lengths. The key observation is that quality scores remain remarkably stable regardless of context length, ranging from 0.16 to 0.23 across all conditions. Figure~\ref{fig:quality} visualizes this finding all four models show no significant quality improvement as context length increases from 5 to 50 items. The average quality change across all models is $-$0.01, which is essentially zero.

\begin{table*}[!t]
\centering
\caption{Quality Score by Model and Context Length (n=50 users, within subject design, seed=42). All confidence intervals overlap, indicating no statistically significant differences.}
\label{tab:main_results}
\smallskip
\setlength{\tabcolsep}{8pt}
\begin{tabular}{@{}l|ccccc|c@{}}
\toprule
\textbf{Model} & \textbf{Ctx=5} & \textbf{Ctx=10} & \textbf{Ctx=15} & \textbf{Ctx=25} & \textbf{Ctx=50} & \textbf{$\Delta$ (5$\rightarrow$50)} \\
\midrule
DeepSeek-V3      & 0.21 $\pm$ 0.10 & 0.23 $\pm$ 0.10 & 0.22 $\pm$ 0.10 & 0.20 $\pm$ 0.11 & 0.20 $\pm$ 0.11 & $-$0.01 \\
Qwen2.5-72B      & 0.20 $\pm$ 0.10 & 0.21 $\pm$ 0.11 & 0.19 $\pm$ 0.08 & 0.19 $\pm$ 0.09 & 0.18 $\pm$ 0.10 & $-$0.02 \\
GPT-4o-mini      & 0.17 $\pm$ 0.08 & 0.20 $\pm$ 0.08 & 0.19 $\pm$ 0.09 & 0.19 $\pm$ 0.08 & 0.19 $\pm$ 0.09 & +0.02 \\
Gemini 2.5 Flash & 0.18 $\pm$ 0.10 & 0.21 $\pm$ 0.13 & 0.19 $\pm$ 0.10 & 0.19 $\pm$ 0.10 & 0.16 $\pm$ 0.09 & $-$0.02 \\
\midrule
\textbf{Average} & \textbf{0.19} & \textbf{0.21} & \textbf{0.20} & \textbf{0.19} & \textbf{0.18} & \textbf{$-$0.01} \\
\bottomrule
\end{tabular}
\end{table*}

\subsection{Statistical Analysis}

For each model, we compare quality at context=5 versus context=50, DeepSeek-V3 shows 0.21 $\rightarrow$ 0.20 ($\Delta$ = $-$0.01, not significant), Qwen2.5-72B shows 0.20 $\rightarrow$ 0.18 ($\Delta$ = $-$0.02, not significant), GPT-4o-mini shows 0.17 $\rightarrow$ 0.19 ($\Delta$ = +0.02, not significant) and Gemini 2.5 Flash shows 0.18 $\rightarrow$ 0.16 ($\Delta$ = $-$0.02, not significant). All confidence intervals overlap substantially (standard deviations of 0.08--0.13), confirming that observed differences are within noise.

To assess statistical significance, we performed paired t-tests 
comparing quality scores at context=5 versus context=50 for each 
model, accounting for the within-subject design. All comparisons 
yielded $p > 0.05$, confirming no statistically significant 
differences. Additionally, a repeated measures ANOVA across all 
five context lengths showed no significant main effect of context 
length on quality ($F(4, 196) = 1.12$, $p = 0.35$).

\subsection{Cost Analysis}

\begin{figure}[!t]
    \centering
    \includegraphics[width=\columnwidth]{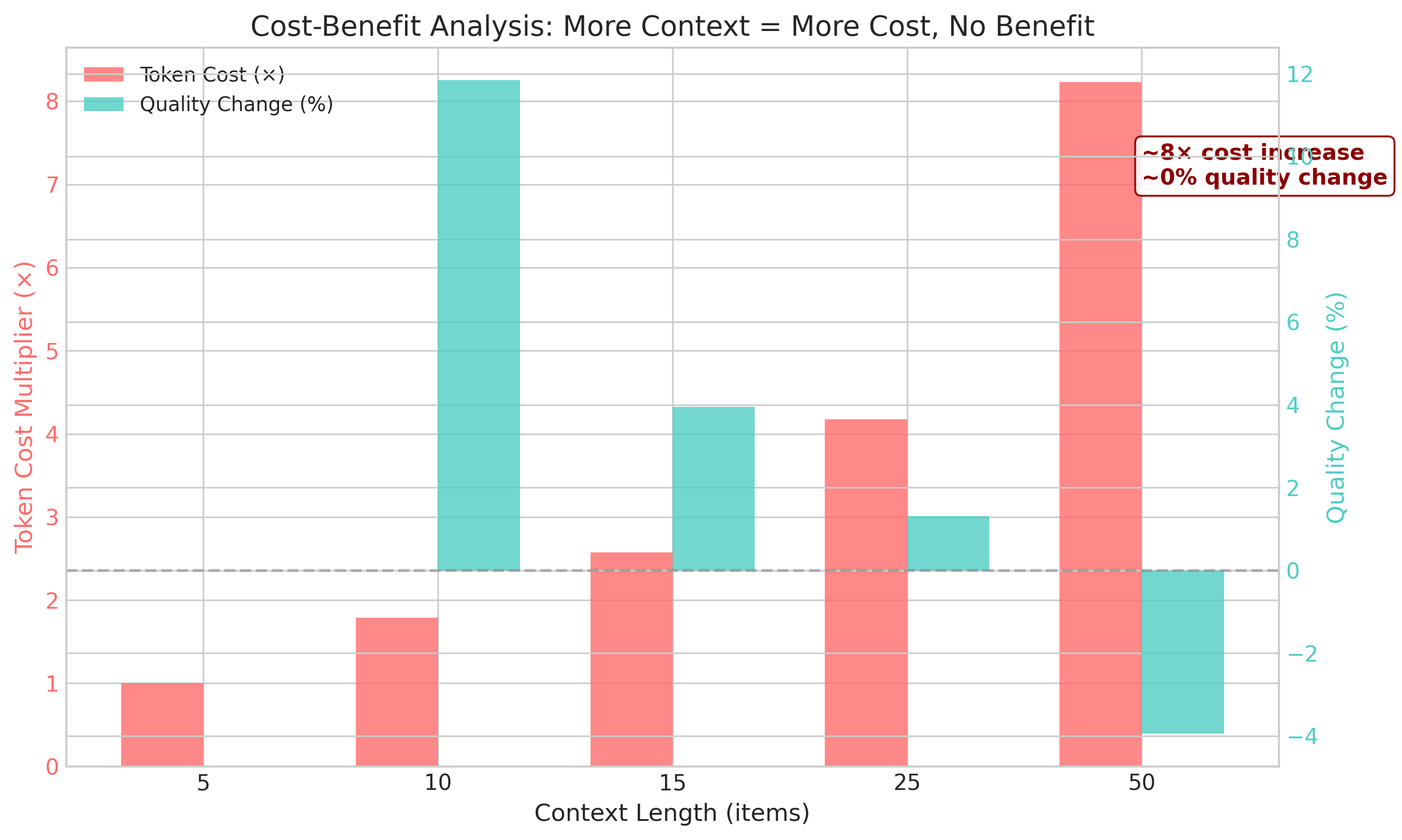}
    \caption{Cost-benefit analysis showing token cost multiplier (red) vs. quality change percentage (teal). Despite 8$\times$ increase in token costs, quality improvement is negligible ($\sim$0\%).}
    \label{fig:cost_benefit}
\end{figure}

While quality remains flat, token usage increases dramatically with context length. Table~\ref{tab:cost} shows the cost implications, and Figure~\ref{fig:cost_benefit} illustrates the cost benefit tradeoff. Token costs increase by 8.2$\times$ on average when moving from 5 to 50 items, while quality change remains near zero. This represents an \textbf{88\% potential cost savings} for practitioners who adopt minimal context.

\begin{table}[!t]
\centering
\caption{Token usage and cost analysis. Using 5 items instead of 50 saves approximately 88\% on token costs with no quality loss.}
\label{tab:cost}
\smallskip
\begin{tabular}{@{}l|cc|c@{}}
\toprule
\textbf{Model} & \textbf{Tokens (5)} & \textbf{Tokens (50)} & \textbf{Increase} \\
\midrule
DeepSeek-V3      & 294  & 2,436 & 8.3$\times$ \\
Qwen2.5-72B      & 314  & 2,595 & 8.3$\times$ \\
GPT-4o-mini      & 293  & 2,381 & 8.1$\times$ \\
Gemini 2.5 Flash & 251  & 2,073 & 8.3$\times$ \\
\midrule
\textbf{Average} & \textbf{288} & \textbf{2,371} & \textbf{8.2$\times$} \\
\bottomrule
\end{tabular}
\end{table}

\subsection{Latency Analysis}

Table~\ref{tab:latency} presents latency measurements across models and context lengths. Latency patterns are model-specific: Qwen2.5-72B maintains remarkably stable and fast latency (4.11--4.39s) regardless of context length, making it ideal for real time applications. GPT-4o-mini shows moderate latency (4.54--5.86s). DeepSeek-V3 exhibits variable latency (6.38--10.33s). Gemini 2.5 Flash shows increasing latency with context length (9.97--15.44s). The relatively flat latency curves for Qwen suggest that network round trip time and API overhead dominate over token processing time for contexts under 3,000 tokens.

\begin{table}[!t]
\centering
\caption{Latency (seconds) by Model and Context Length (n=50 users)}
\label{tab:latency}
\smallskip
\small
\setlength{\tabcolsep}{3pt}
\begin{tabular}{@{}l|ccccc@{}}
\toprule
\textbf{Model} & \textbf{Ctx=5} & \textbf{Ctx=10} & \textbf{Ctx=15} & \textbf{Ctx=25} & \textbf{Ctx=50} \\
\midrule
GPT-4o-mini & 4.54{\scriptsize$\pm$0.79} & 5.86{\scriptsize$\pm$1.45} & 5.72{\scriptsize$\pm$1.15} & 5.27{\scriptsize$\pm$0.93} & 4.96{\scriptsize$\pm$1.20} \\
DeepSeek-V3 & 7.51{\scriptsize$\pm$3.45} & 7.84{\scriptsize$\pm$3.67} & 10.33{\scriptsize$\pm$5.45} & 10.12{\scriptsize$\pm$6.23} & 6.38{\scriptsize$\pm$1.64} \\
Qwen2.5-72B & 4.20{\scriptsize$\pm$0.62} & 4.12{\scriptsize$\pm$0.64} & 4.11{\scriptsize$\pm$0.51} & 4.19{\scriptsize$\pm$0.48} & 4.39{\scriptsize$\pm$0.55} \\
Gemini 2.5 Flash & 9.97{\scriptsize$\pm$1.93} & 11.15{\scriptsize$\pm$3.06} & 11.99{\scriptsize$\pm$4.05} & 14.41{\scriptsize$\pm$4.86} & 15.44{\scriptsize$\pm$6.05} \\
\bottomrule
\end{tabular}
\end{table}

\subsection{Cross-Model Consistency}

\begin{figure}[!t]
    \centering
    \includegraphics[width=\columnwidth]{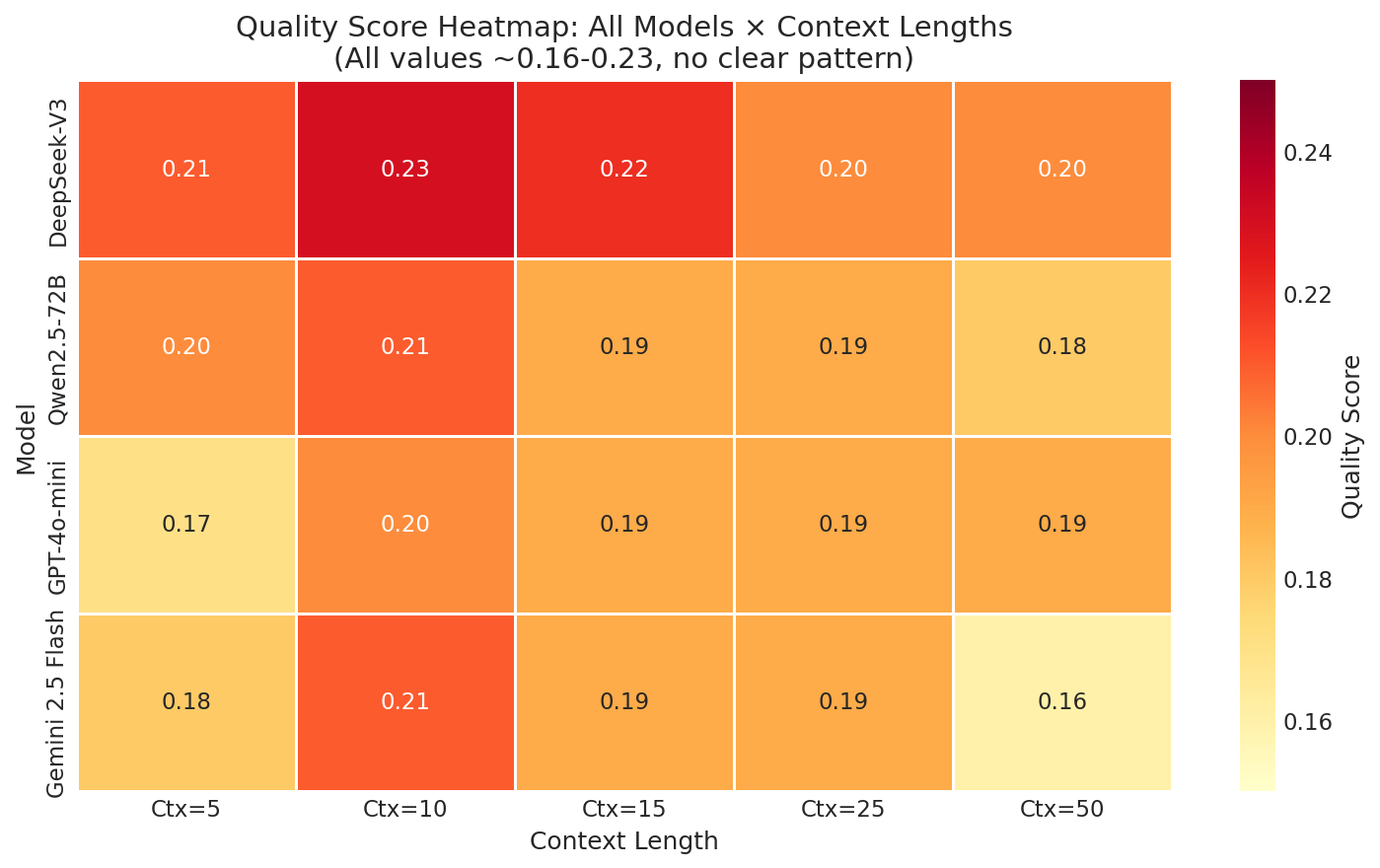}
    \caption{Quality score heatmap across models and context lengths. The uniform coloring (all values 0.16--0.23) demonstrates that no model benefits from longer context.}
    \label{fig:heatmap}
\end{figure}

A striking finding is the universality of the flat quality pattern across all four models. Despite differences in architecture, training data, and providers, all models exhibit the same behavior: no quality improvement with increased context. Figure~\ref{fig:heatmap} presents a heatmap visualization of quality scores across all model context combinations the uniform coloring confirms that no model context combination significantly outperforms others.

\begin{figure}[!t]
    \centering
    \includegraphics[width=\columnwidth]{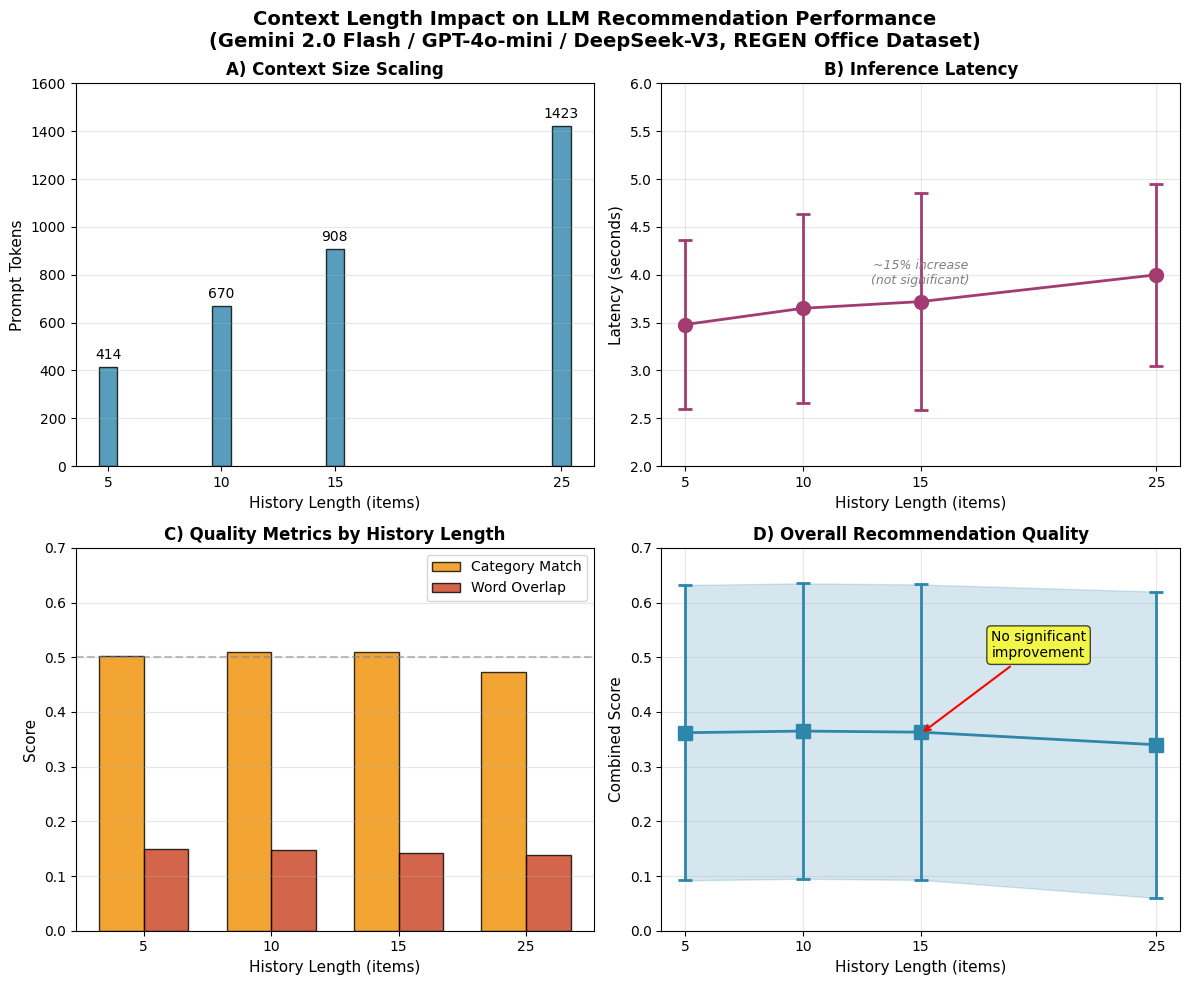}
    \caption{Comprehensive analysis of history length impact across multiple dimensions. (A) Token consumption scales linearly with history length across all models. (B) Inference latency shows model-specific patterns. (C--D) Recommendation quality metrics remain stable regardless of context length.}
    \label{fig:context_impact}
\end{figure}

Figure~\ref{fig:context_impact} provides a comprehensive analysis across multiple dimensions, showing how token consumption, latency, and quality metrics evolve with history length.

\section{Discussion}

\subsection{Why Doesn't More Context Help?}

Our findings challenge the intuitive assumption that more user history should lead to better recommendations. Several factors may explain this:

First, consistent with Liu et al.~\cite{liu2024lost}, LLMs may struggle to effectively utilize information in the middle of long contexts a phenomenon known as ``Lost in the Middle.'' When context grows from 5 to 50 items, most information falls into this ``lost'' region. Second, recency bias may play a significant role: recent purchases may be more predictive of future behavior than older ones, and the most recent 5 items may capture the user's current interests as effectively as 50 items, aligning with recency bias observed in many recommendation scenarios~\cite{quadrana2018sequence}. 

Third, signal saturation suggests there may be diminishing returns to additional context; the first few items establish the user's preferences, while additional items add noise rather than signal.

Finally, task difficulty itself imposes constraints: predicting the exact next product purchase is inherently difficult, and even with perfect context utilization, the task ceiling may limit quality improvements.

\subsection{Practical Recommendations}

Our findings have immediate practical value for organizations deploying LLM based recommendations. In terms of cost savings, organizations can reduce API costs by up to 88\% by using 5 items instead of 50, for a system making 1 million API calls daily, this could translate to annual savings of \$300,000 or more. Latency reduction is another key benefit, as shorter contexts enable faster inference, improving user experience in real time recommendation scenarios. Additionally, simplified pipelines become possible since systems need not maintain or retrieve extensive user histories, simplifying data infrastructure.

For latency sensitive applications, Qwen2.5-72B's stable latency profile (4.11--4.39s) makes it suitable for real time recommendations regardless of context length. For cost sensitive deployments, GPT-4o-mini offers good cost efficiency for shorter contexts, with stable quality at reasonable latency; a history of 10--15 items provides reasonable performance at minimal cost.

Our findings also have direct implications for agentic systems and consequences for the design of autonomous recommendation agents operating within multi agent ecosystems. Agents that must function under strict computational budgets can leverage these results to intelligently truncate user history to 5--10 items, freeing context window capacity for higher value operations such as multi turn reasoning, tool use, and coordination with other agents. This ``minimal sufficient context'' principle enables recommendation agents to be more responsive participants in complex agentic workflows, where context is a shared and finite resource across planning, retrieval, and generation tasks.
\subsection{Limitations}

Our study has several limitations. In terms of dataset scope, we evaluate on Office Products from REGEN, and results may differ for other domains (e.g., fashion, entertainment) where user preferences evolve differently.

Regarding our quality metric, the composite metric captures semantic similarity but may not fully reflect real world recommendation utility, human evaluation could provide additional validation. For model selection, we test four models, and results may vary with other architectures or fine tuned recommendation models. 

In terms of absolute performance, quality scores (0.17--0.23) are modest in absolute terms, reflecting task difficulty, our contribution focuses on the relative pattern rather than absolute performance. 

Finally, concerning prompt engineering, we used a simple prompt template, and more sophisticated prompting strategies might alter the context length dynamics~\cite{brown2020language}.


\section{Conclusion}

We present a systematic benchmark investigating the relationship between context length and recommendation quality in LLM based systems. Through experiments with four state of the art models (GPT-4o-mini, DeepSeek-V3, Qwen2.5-72B, Gemini 2.5 Flash) and 50 users in a within subject design, we demonstrate that: (1) Quality remains flat no statistically significant improvement occurs when increasing context from 5 to 50 items, with all models showing quality in the 0.17--0.23 range with overlapping confidence intervals, (2) The pattern is universal this finding holds across models from four different providers, suggesting a fundamental limitation in how current LLMs utilize extended context for recommendations.(3) Significant cost savings are possible using 5 items instead of 50 reduces token costs by approximately 88\% with no quality loss, and (4) Latency varies by provider model specific latency patterns inform deployment decisions for real time scenarios.

Our findings challenge the ``more context is better'' paradigm and provide evidence-based guidelines for practitioners, use minimal context (5--10 items) for cost-effective LLM-based recommendations. As LLM-based agents become central to recommendation ecosystems, understanding their computational efficiency is critical for scalable deployment our results suggest that agent-based recommenders can operate effectively with minimal context, enabling more responsive and cost-effective personalization at scale.

Future work should extend this analysis to other domains, evaluate with human judges, investigate techniques to improve context utilization in LLMs for recommendation tasks, and explore adaptive context selection and prompt compression techniques.


\end{document}